\documentclass[preprint,superscriptaddress,showpacs,preprintnumbers,amsmath,amssymb]{revtex4}

\usepackage{graphicx}
\usepackage{dcolumn}
\usepackage{bm}

\def\lesssim{\ \raise.3ex\hbox{$<$}\kern-0.8em\lower.7ex\hbox{$\sim$}\ }
\def\gesim{\ \raise.3ex\hbox{$>$}\kern-0.8em\lower.7ex\hbox{$\sim$}\ }

\def\rnum#1{\expandafter{\romannumeral #1}} 
\def\Rnum#1{\uppercase\expandafter{\romannumeral #1}} 
\begin{document}
\title{Coexistence of superfluid gap and pseudogap in the BCS-BEC crossover regime of a trapped Fermi gas below $T_{\rm c}$}
\author{Ryota Watanabe}
\email{rwatanab@rk.phys.keio.ac.jp}
\affiliation{Faculty of Science and Technology, Keio University,
3-14-1 Hiyoshi, Kohoku-ku, Yokohama 223-8522, Japan}

\author{Shunji Tsuchiya}
\affiliation{Department of Physics, Faculty of Science, Tokyo University
of Science, 1-3 Kagurazaka, Shinjuku-ku, Tokyo 162-8601, Japan}
\affiliation{Research and Education Center for Natural Sciences, Keio University, Yokohama, Japan, 4-1-1 Hiyoshi, Kanagawa 223-8521, Japan}

\author{Yoji Ohashi}
\affiliation{Faculty of Science and Technology, Keio University,
3-14-1 Hiyoshi, Kohoku-ku, Yokohama 223-8522, Japan}

\date{\today}       
\begin{abstract}
We investigate strong pairing fluctuations and effects of a harmonic trap in the superfluid phase of an ultracold Fermi gas.
Including amplitude and phase fluctuations of the inhomogeneous superfluid order parameter $\Delta(r)$ in a trap within a combined $T$-matrix theory with the local density approximation, we examine local properties of single-particle excitations and a thermodynamic quantity in the BCS-BEC crossover region.
Below the superfluid phase transition temperature $T_{\rm c}$, we show that inhomogeneous pairing fluctuations lead to a shell structure of the gas cloud in which the spatial region where the ordinary BCS-type superfluid density of states appears is surrounded by the region where the pseudogap associated with strong pairing fluctuations dominates single-particle excitations. The former spatial region enlarges to eventually cover the whole gas cloud far below $T_{\rm c}$. We also examine how this shell structure affects the photoemission spectrum, as well as the local pressure. Since a cold Fermi gas is always trapped in a harmonic potential, our results would be useful for the study of strong-coupling superfluid physics, including this realistic situation.
\end{abstract}
\maketitle

\section{Introduction}
\label{section1}
Since the background physics of superfluid Fermi gases\cite{Regal,Zwierlein,Kinast,Bartenstein} is similar to that of metallic superconductivity, the former system is now expected as a useful quantum simulator for the latter. A tunable pairing interaction associated with a Feshbach resonance\cite{Timmermans,Holland,OHASHI2,GIORGINI,BLOCH,KETTERLE} in a cold Fermi gas enables us to study Fermi superfluids from the weak-coupling BCS (Bardeen-Cooper-Schrieffer) regime to the strong-coupling BEC (Bose-Einstein condensation) regime in a unified manner\cite{Timmermans,Holland,OHASHI2,GIORGINI,BLOCH,KETTERLE,EAGLES,LEGGETT,NSR,SADEMELO,Randeria2}. In this BCS-BEC crossover, one can systematically examine strong-coupling effects by adjusting the interaction strength. 
\par
While there exist various similarities between a superfluid Fermi gas and metallic superconductivity, the presence of a trap potential is peculiar to the former. Because of this confined geometry, physical quantities naturally become inhomogeneous. Thus, when an experiment has no spatial resolution, it gives spatially averaged data. For example, the photoemission-type experiment developed by JILA group\cite{STEWART,GAEBLER,KOHL,SOMMER} so far has no spatial resolution, so that the observed data involve single-particle excitation spectra at various spatial positions. Thus, the spatial inhomogeneity is a crucial key in considering single-particle properties of a superfluid Fermi gas by using this experiment. Since a bulk superconductor is usually a uniform system, this problem is also important in using a superfluid Fermi gas as a quantum simulator for superconductivity. 
\par
In this paper, we investigate effects of a harmonic trap on strong-coupling properties of a superfluid Fermi gas in the BCS-BEC crossover region. In particular, as a typical strong-coupling phenomenon, we deal with the pseudogap problem\cite{Perali,Pieri1,Pieri2,Haussmann,TSUCHIYA1,TSUCHIYA2,TSUCHIYA3,Hu}. In this phenomenon, strong pairing fluctuations induce a dip structure in the density of states above the superfluid phase transition temperature $T_{\rm c}$. In our previous paper for a uniform Fermi gas\cite{WATANABE1}, we showed that the pseudogap still remains just below $T_{\rm c}$. Then, in a trapped superfluid Fermi gas, we can expect the inhomogeneous situation that while the BCS-type superfluid density of states appears in the trap center, the pseudogap is still dominant in the outer region of the gas cloud (where the superfluid order parameter is small and pairing fluctuations are strong). 
\par
To confirm such a shell structure, it is convenient to examine the superfluid local density of states (LDOS), as well as the local spectral weight (LSW). In this paper, using a combined $T$-matrix theory with the local density approximation (LDA), we identify the pseudogap region in the phase diagram with respect to the interaction strength, temperature, and spatial position. We also discuss how the shell structure affects the photoemission spectrum.
\par
Besides single-particle excitations, thermodynamic properties has also attracted much attention in the BCS-BEC crossover region\cite{GIORGINI,BLOCH,Haussmann2,Haussmann3,Hu2,Ohara,Luo,Ho2,Horikoshi,Nascimbene}. In this paper, we consider the local pressure $P(r)$ as a typical thermodynamic quantity. Recently, Ho and Zhou\cite{Ho2} proposed a useful idea to determine $P(r)$ from the density profile. Using this, Nascimbene and co-workers\cite{Nascimbene} measured $P(r)$ in the unitarity limit of a $^6$Li Fermi gas, as a function of the LDA fugacity $\zeta(r)=e^{\mu(r)/k_{\rm B}T}$ (where $\mu(r)$ is the LDA chemical potential). They reported that the observed pressure is well described by the Fermi liquid theory. In this paper, we clarify to what extent strong pairing fluctuations affect $P(r)$, comparing our results with the experimental data\cite{Nascimbene}.
\par
This paper is organized as follows. In Sec.II, we explain our formulation. In Sec.III, we present our numerical results on LDOS and LSW in the BCS-BEC crossover below $T_{\rm c}$. We also identify the region where the pseudogap dominates single-particle excitations in the phase diagram of a trapped Fermi gas. In Sec.IV, we consider the photoemission spectrum. We clarify how the inhomogeneous pseudogap affects this quantity below $T_{\rm c}$. In Sec.V, we treat the local pressure $P(r)$. We show that the calculated $P(r)$ agrees well with the recent experiment on a $^6{\rm Li}$ Fermi gas\cite{Nascimbene}. Throughout this paper, we set $\hbar=k_{\rm B}=1$.
\par

\section{Formulation}
We consider a two-component superfluid Fermi gas in a harmonic potential, described by the BCS Hamiltonian. In the Nambu representation, it has the form\cite{WATANABE1,OHASHI1},
\begin{equation}
H=\sum_{\bm p}\Psi_{\bm p}^\dagger
\Bigl[
\xi_p\tau_3-\Delta\tau_1
\Bigr]\Psi_{\bm p}
-U\sum_{\bm q}\rho_+({\bm q})\rho_-(-{\bm q}).
\label{Hamiltonian}
\end{equation}
(Effects of a harmonic trap is later included within LDA.) Here, 
\begin{eqnarray}
\Psi_{\bm p}=
\left(
\begin{array}{c}
c_{\bm p\uparrow}\\
c_{-\bf p\downarrow}^\dagger
\end{array}
\right)
\label{Nambu}
\end{eqnarray}
is the two-component Nambu field, where $c_{\bm p,\sigma}^\dagger$ is the creation operator of a Fermi atom with pseudospin $\sigma$ ($=\uparrow,\downarrow$), describing two atomic hyperfine states. $\xi_{\bm p}=\varepsilon_{\bm p}-\mu={\bm p}^2/(2m)-\mu$ is the kinetic energy, measured from the Fermi chemical potential $\mu$ (where $m$ is an atomic mass). $-U$ ($<0$) is a pairing interaction, which is assumed to be tunable by adjusting the threshold energy of a Feshbach resonance. $\tau_j$ ($j=1,2,3$) are the Pauli matrices, acting on the particle-hole space. In this paper, we take the superfluid order parameter, 
\begin{equation}
\Delta=U\sum_{\bm p}\langle c_{-{\bm p}\downarrow}c_{{\bm p}\uparrow} \rangle,
\label{delta}
\end{equation}
to be real and be parallel to the $\tau_1$-component. In this case, the generalized density operators $\rho_j({\bm q})=\sum_{\bm p}\Psi_{\bm {p+q/2}}^\dagger\tau_j\Psi_{\bm {p-q/2}}$ ($j=1,2$) in $\rho_{\pm}({\bm q})=[\rho_1({\bm q})\pm i\rho_2({\bm q})]/2$ describe amplitude and phase fluctuations of the superfluid order parameter, respectively\cite{noteZ}. 
\par
\begin{figure}[t]
\includegraphics[width=8cm]{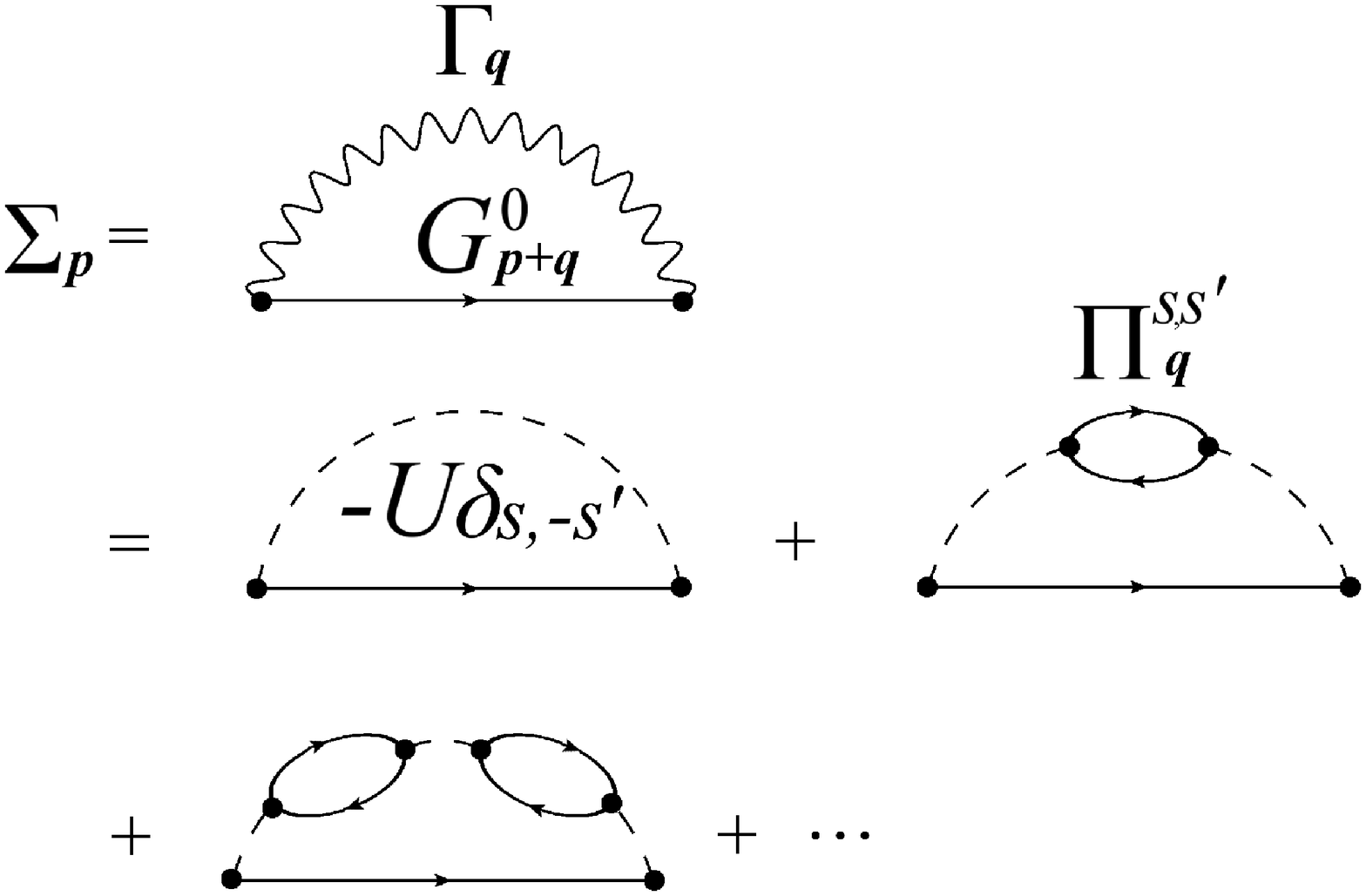}
\caption{(a) Self-energy correction $\Sigma_{\bm p}(i\omega_n,r)$ in the combined $T$-matrix theory with LDA. The wavy line is the particle-particle scattering matrix $\Gamma^{s,s'}_{\bm q}(i\nu_n,r)$ ($s,s'=\pm$), describing amplitude and phase fluctuations of the superfluid order parameter. The solid and dashed lines describe the LDA mean-field Green's function $G_{\bm p}^0(i\omega_n,r)=[i\omega_n-\xi_{\bm p}(r)\tau_3+\Delta(r)\tau_1]^{-1}$, and the pairing interaction $-U$, respectively. The bubble diagram describes the pair-correlation function $\Pi_{\bm q}^{s,s'}(i\nu_n,r)$.}
\label{fig1}
\end{figure}
\par
As usual, we measure the interaction strength in terms of the $s$-wave scattering length $a_s$, which is related to the pairing interaction $-U$ as\cite{Randeria2},
\begin{equation}
{4\pi a_s \over m}=-
{U \over 1-U\sum_{\bm p}^{\omega_{\rm c}}{1 \over 2\varepsilon_{\bm p}}},
\end{equation}
where $\omega_{\rm c}$ is a high-energy cutoff.
\par
We now include effects of a harmonic trap potential $V(r)=m\omega_{\rm tr}^2r^2/2$ within LDA. This extension is achieved by simply replacing the Fermi chemical potential $\mu$ by the LDA expression $\mu(r)=\mu-V(r)$\cite{TSUCHIYA2,OHASHI3}. Various quantities then depend on the spatial position $r$ through $\mu(r)$. For example, the $2\times2$-matrix LDA single-particle thermal Green's function is given by
\begin{equation}
G_{\bm p}(i\omega_n,r)=\frac{1}{i\omega_n-\xi_{\bm p}(r)\tau_3+\Delta(r)\tau_1-\Sigma_{\bm p}(i\omega_n,r)},
\end{equation}
where $\xi_{\bm p}(r)=\varepsilon_{\bm p}-\mu(r)$, and $\omega_n$ is the fermion Matsubara frequency. $\Delta(r)$ is the LDA superfluid order parameter. The LDA self-energy $\Sigma_{\bm p}(i\omega_n,r)$ involves effects of pairing fluctuations within the $T$-matrix approximation, which is diagrammatically given by Fig. \ref{fig1}\cite{Pieri2,WATANABE1}. Summing up these diagrams, we obtain
\begin{equation}
\Sigma_{\bm p}(i\omega_n,r)=-T\sum_{{\bm q},\nu_n}\sum_{s,s^\prime=\pm}\Gamma^{s,s^\prime}_{\bm q}(i\nu_n,r)\tau_{-s}G^0_{{\bm p}+{\bm q}}(i\omega_n+i\nu_n,r)\tau_{-s^\prime}.
\end{equation}
Here, $\tau_\pm=[\tau_1\pm i\tau_2]/2$, and $\nu_n$ is the boson Matsubara frequency. $G^0_{\bm p}(i\omega,r)=[i\omega_n-\xi_{\bm p}(r)\tau_3+\Delta(r)\tau_1]^{-1}$ is the LDA mean-field Green's function. The particle-particle scattering matrix $\Gamma^{s,s^\prime}_{\bm q}(i\nu_n,r)$ has the form
\begin{equation}
\left(
\begin{array}{cc}
\Gamma^{+-}_{\bm q}(i\nu_n,r)&\Gamma^{++}_{\bm q}(i\nu_n,r)\\
\Gamma^{--}_{\bm q}(i\nu_n,r)&\Gamma^{-+}_{\bm q}(i\nu_n,r)
\end{array}
\right)=-U\left[1+U\left(
\begin{array}{cc}
\Pi^{-+}_{\bm q}(i\nu_n,r)&\Pi^{++}_{\bm q}(i\nu_n,r)\\
\Pi^{--}_{\bm q}(i\nu_n,r)&\Pi^{+-}_{\bm q}(i\nu_n,r)
\end{array}
\right)
\right]^{-1},
\end{equation}
where
\begin{eqnarray}
\Pi^{s,s^\prime}_{\bm q}(i\nu_n,r)
&=&T\sum_{\bm p,\omega_n}{\rm Tr}\left[\tau_sG^0_{{\bm p}+{\bm q}/2}(i\omega_n+i\nu_n,r)\tau_{s'}G^0_{{\bm p}-{\bm q}/2}(i\omega_n,r)\right]
\label{eq.paipp}
\end{eqnarray}
is the lowest-order pair correlation function with respect to the pairing interaction. Carrying out the $\omega_n$-summation in Eq. (\ref{eq.paipp}), we obtain
\begin{eqnarray}
\Pi_{\bm q}^{++}(i\nu_n,r)
&=&
\frac{1}{4}\sum_{s=\pm1}\sum_{\bm
 p}\frac{s\Delta(r)^2}
 {E_{{\bm p}+{\bm q}/2}(r)E_{{\bm p}-{\bm q}/2}(r)}
 \frac{E_{{\bm p}+{\bm q}/2}(r)+sE_{{\bm p}-{\bm q}/2}(r)}
 {\nu_n^2+(E_{{\bm p}+{\bm q}/2}(r)+sE_{{\bm p}-{\bm q}/2}(r))^2}
\nonumber\\ 
&&\qquad\times\left[\tanh\left({E_{{\bm p}+{\bm q}/2}(r) \over 2T}\right)+s\tanh\left({E_{{\bm p}-{\bm q}/2}(r) \over 2T}\right)\right],
\label{explicitpi1}
\end{eqnarray}
\begin{eqnarray}
\Pi_{\bm q}^{+-}(i\nu_n,r)&=&\frac{1}{4}\sum_{s=\pm1}\sum_{\bm
 p}\left[\left[1+s\frac{\xi_{{\bm p}+{\bm q}/2}(r)\xi_{{\bm
		     p}-{\bm q}/2}(r)}{E_{{\bm p}+{\bm q}/2}(r)E_{{\bm
		     p}-{\bm q}/2}(r)}\right]\frac{1}{i\nu_n-(E_{{\bm
 p}+{\bm q}/2}(r)+sE_{{\bm p}-{\bm q}/2}(r))}\right.\nonumber\\ 
&&+\left.\left[1-\frac{\xi_{\bm p+\bf q/2}(r)}{E_{{\bm p}+{\bm
	  q}/2}(r)}\right]\left[1-s\frac{\xi_{{\bm p}-{\bm q}/2}(r)}{E_{{\bm
	  p}-{\bm q}/2}(r)}\right]\frac{i\nu_n}{\nu_n^2+(E_{{\bm p}+{\bm q}/2}(r)+sE_{{\bm
	  p}-{\bm q}/2}(r))^2} \right]\nonumber\\ 
&&\qquad\times\left[\tanh\left({E_{{\bm p}+{\bm q}/2}(r) \over 2T}\right)+s\tanh\left({E_{{\bm p}-{\bm q}/2}(r) \over 2T}\right)\right].
\label{explicitpi2}
\end{eqnarray}
The other components are given by $\Pi_{\bm q}^{--}(i\nu_n,r)=\Pi_{\bm q}^{++}(i\nu_n,r)$, and $\Pi_{\bm q}^{-+}(i\nu_n,r)=\Pi_{\bm q}^{+-}(-i\nu_n,r)$. In Eqs. (\ref{explicitpi1}) and (\ref{explicitpi2}), $E_{\bm p}(r)=\sqrt{\xi_{\bm p}(r)^2+\Delta(r)^2}$ is the LDA Bogoliubov single-particle excitation spectrum. 
\par
\begin{figure}[t]
\includegraphics[width=10cm]{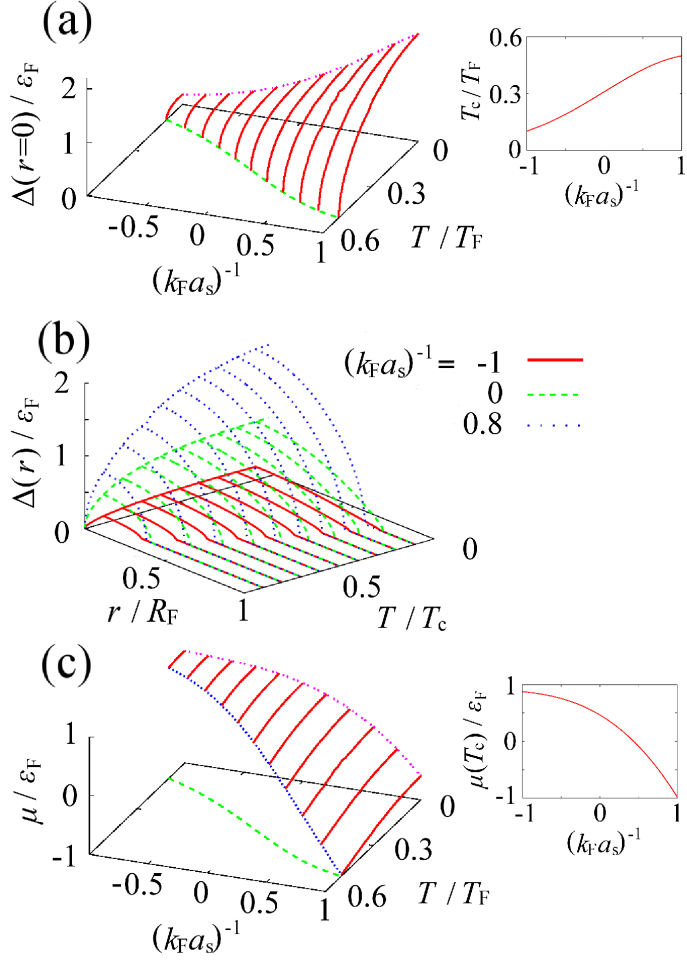}
\caption{(Color online) (a) Calculated LDA superfluid order parameter $\Delta(r=0)$. The inset shows $T_{\rm c}$, normalized by the Fermi temperature $T_{\rm F}$. (b) Spatial variation of $\Delta(r)$. $R_{\rm F}=\sqrt{2\varepsilon_{\rm F}/(m\omega_{\rm tr}^2)}$ is the Thomas Fermi radius, where $\varepsilon_{\rm F}$ is the Fermi energy. (c) Chemical potential $\mu$. The dashed line shows $T_{\rm c}$. The inset shows $\mu(T=T_{\rm c})$.
}
\label{fig2}
\end{figure}
\par
We self-consistently determine the local superfluid order parameter $\Delta(r)$ and the chemical potential $\mu$, by solving the LDA gap equation,
\begin{equation}
1=U\sum_{\bm p}\frac{1}{2E_{\bm p}(r)}\tanh{E_{\bm p}(r) \over 2T},
\label{eq.10}
\end{equation}
together with the equation for the total number $N$ of Fermi atoms,
\begin{equation}
N=\int_0^\infty 4\pi r^2dr n(r),
\label{eq.10b}
\end{equation}
where $n(r)=2T\sum_{{\bm p},\omega_n}G_{\bm p}(i\omega_n,r)|_{11}e^{i\delta\omega_n}$ is the particle density. Within the framework of LDA, $T_{\rm c}$ is determined from the BCS-type $T_{\rm c}$-equation in the trap center ($r=0$)\cite{TSUCHIYA2,TSUCHIYA3,OHASHI3},
\begin{equation}
1=U\sum_{\bm p}\frac{1}{2\xi_{\bm p}}\tanh{\xi_{\bm p} \over 2T}.
\label{eq.11}
\end{equation}
\par
Figure \ref{fig2} shows the self-consistent solutions of $\Delta(r)$ and $\mu$.  We will use these numerical results in calculating various quantities in later sections.

\par
\begin{figure}[t]
\includegraphics[width=8cm]{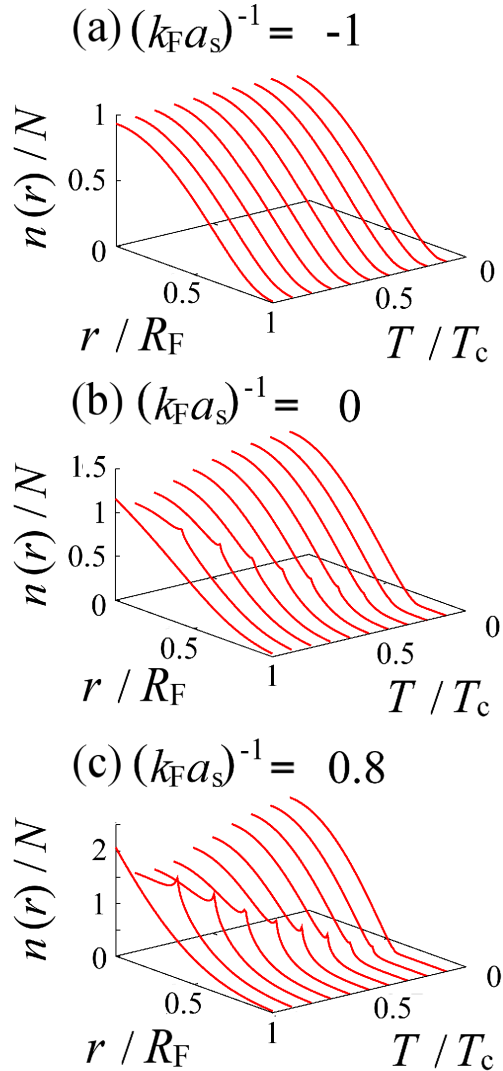}
\caption{(Color online) Atomic density profile $n(r)$ in the BCS-BEC crossover below $T_{\rm c}$. (a) $(k_{\rm F}a_{\rm s})^{-1}=-1$ (BCS side). (b) $(k_{\rm F}a_{\rm s})^{-1}=0$ (unitarity limit). (c) $(k_{\rm F}a_{\rm s})^{-1}=0.8$ (BEC side).}
\label{fig3}
\end{figure}

\begin{figure}[t]
\includegraphics[width=6cm]{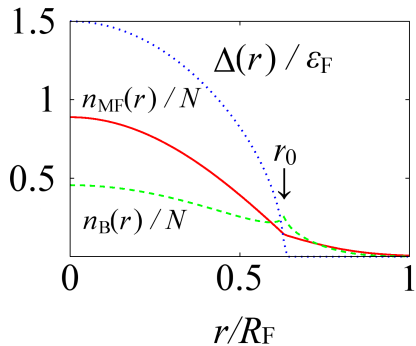}
\caption{(Color online) Mean-filed contribution $n_{\rm MF}(r)$ and fluctuation contribution $n_{\rm B}(r)$ to the density profile $n(r)$ at $0.5T_{\rm c}$ in the unitarity limit. The LDA superfluid order parameter $\Delta(r_0)$ vanishes at $r=r_0$, at which a cusp appears in $n_{\rm B}(r)$, .}
\label{fig4}
\end{figure}
\par
We note that the LDA gap equation (\ref{eq.10}) gives the vanishing superfluid order parameter $\Delta(r)=0$ for $r\ge r_0$, where $r_{\rm 0}$ is determined from the equation,
\begin{equation}
1=U\sum_{\bm p}\frac{1}{2\xi_{\bm p}(r_0)}\tanh{\xi_{\bm p}(r_0) \over 2T}.
\label{eq.gap2}
\end{equation}
However, this vanishing $\Delta(r>r_0)$ is an artifact of LDA, because the superfluid order parameter should be finite everywhere in the gas cloud below $T_{\rm c}$. Since the correct $\Delta(r)$ would be large around the trap center, $r_0$ obtained in LDA should be interpreted as a characteristic radius, inside of which the magnitude of $\Delta(r)$ is large. 
\par
As pointed out in Ref.\cite{Pieri1}, the LDA atomic density profile $n(r)$ has a cusp at $r=r_0$, which becomes more remarkable as one approaches the strong-coupling regime, as shown in Fig. \ref{fig3}. Dividing $n(r)$ into the sum of the mean-field part $n_{\rm MF}(r)=2T\sum_{{\bm p},\omega_n}G_{\bm p}^0(\omega_n,r)|_{11}e^{i\delta\omega_n}$ and the fluctuation contribution, 
\begin{equation}
n_{\rm B}(r)=2T\sum_{{\bm p},\omega_n}\left[G_{\bm p}(\omega_n,r)-G_{\bm p}^0(\omega_n,r)\right]|_{11}e^{i\delta\omega_n}.
\end{equation}
we find in Fig. \ref{fig4} that the cusp only appears in $n_{\rm B}(r)$. In a Bose gas BEC, the LDA density profile at $T_{\rm c}$ is known to exhibit a cusp at $r=0$\cite{Pethick}, which is also seen in Figs. \ref{fig3}(b) and (c) at $T_{\rm c}$. Since $n_{\rm B}(r)$ reduces to twice the molecular density profile in the BEC limit, the cusp singularity seen in Fig. \ref{fig3} is found to be the same artifact of LDA as in the Bose gas case\cite{Pethick}.
\par
Once $\Delta(r)$ and $\mu$ are determined, the LDA single-particle spectral weight (LSW) $A(\bf p,\omega,\bf r)$ is conveniently calculated from the analytic continued Green's function as
\begin{equation}
A({\bm p},\omega,r)=-\frac{1}{\pi}{\rm Im}
G_{\bm p}(i\omega_n\to\omega+i\delta,r)|_{11}.
\label{SW}
\end{equation}
The local density of states (LDOS) $\rho(\omega,r)$ is related to LSW as
\begin{equation}
\rho(\omega,r)=\sum_{\bm p}A({\bm p},\omega,r).
\label{DOS}
\end{equation}
\par
For the infinitesimally small positive number $\delta$ appearing in Eq. \ref{SW}, to avoid unphysical fine structures in LDOS and LSW, we take $\delta=0.01\varepsilon_{\rm F}$ in numerical calculations.
\par
The local spectral weight (LSW) is also related to the photoemission spectrum\cite{STEWART,GAEBLER}. In this experiment, atoms in one of the two atomic hyperfine states $(\equiv |\uparrow\rangle$) are transferred to another hyperfine state $|3\rangle$ ($\ne |\uparrow\rangle, |\downarrow\rangle$) by rf-pulse, and one measures the rf-tunneling current from $|\uparrow\rangle$ to $|3\rangle$. In $^{40}$K Fermi gases, because the so-called final state interaction can be safely ignored\cite{STEWART,GAEBLER}, $|3\rangle$ may be treated as a non-interacting state. Using the linear response theory, we obtain the rf-tunneling current as,
\begin{equation}
I({\bm p},\Omega,r)=2\pi t_{\rm F}^2A_{\bm p}(\xi_{\bm p}(r)-\Omega,r)f(\xi_{\bm p}(r)-\Omega).
\label{eq.100}
\end{equation}
For the derivation of Eq. (\ref{eq.100}), we refer to Ref.\cite{TSUCHIYA3}. $t_{\rm F}$ represents a transfer matrix element between $|\uparrow\rangle$ and $|3\rangle$. $f(\varepsilon)$ is the Fermi distribution function. Noting that the current photoemission-type experiment does not have spatial resolution, we take the spatial average of Eq. (\ref{eq.100}) as
\begin{equation}
I_{\rm ave}({\bm p},\Omega)=\frac{2\pi t_{\rm F}^2}{4\pi R_{\rm F}^3/3}\int d{\bm r}A_{\bm p}(\xi_{\bm p}(r)-\Omega,r)f(\xi_{\bm p}(r)-\Omega).
\label{eq.101}
\end{equation}
Here, $R_{\rm F}=\sqrt{2\varepsilon_{\rm F}/(m\omega_{\rm tr}^2)}$ is the Thomas-Fermi radius (where $\varepsilon_{\rm F}$ is the Fermi energy), and $\Omega=\omega_{\rm L}-\omega_3$ is the energy difference between the incident photon energy $\omega_{\rm L}$ and the energy $\omega_3$ of the final state $|3\rangle$. Equation (\ref{eq.101}) is related to the observed photoemission spectrum\cite{STEWART,GAEBLER} $p^2\overline{A_{\bm p}(\omega)f(\omega)}$, as well as the occupied density of states $\overline{\rho(\omega)f(\omega)}$, as
\begin{eqnarray}
p^2\overline{A_{\bm p}(\omega)f(\omega)}&=&p^2 I_{\rm ave}({\bm p},\Omega\to \xi_p-\omega),
\label{eq.102}
\\
\overline{\rho(\omega)f(\omega)}
&=&
{1 \over 2\pi^2}
\int dp p^2\overline{A_{\bm p}(\omega)f(\omega)}.
\label{eq.103}
\end{eqnarray}
\par
To calculate the local pressure in a gas cloud, we employ the idea proposed by Ho and Zhou\cite{Ho2}. That is, using the Gibbs-Duham equation for the local pressure, $dP(r)=n(r)\mu(r)+s(r)dT$ (where $s(r)$ is the entropy density), we obtain $dP(r)=n(r)d\mu(r)$ for a fixed value of $T$. Then, using the relation $d\mu(r)=m\omega^2rdr$ and $P(r\to\infty)=0$, one finds\cite{Ho2}
\begin{equation}
P(r)=m\omega^2\int_{\infty}^rr^\prime dr^\prime n(r^\prime).
\label{eq.P}
\end{equation}
Substituting the calculated particle density $n(r)$ shown in Fig. \ref{fig3} into Eq. (\ref{eq.P}), we numerically carry out the integration in Eq. (\ref{eq.P}) to obtain the local pressure $P(r)$. 
\par
\begin{figure}[t]
\includegraphics[width=.9\textwidth]{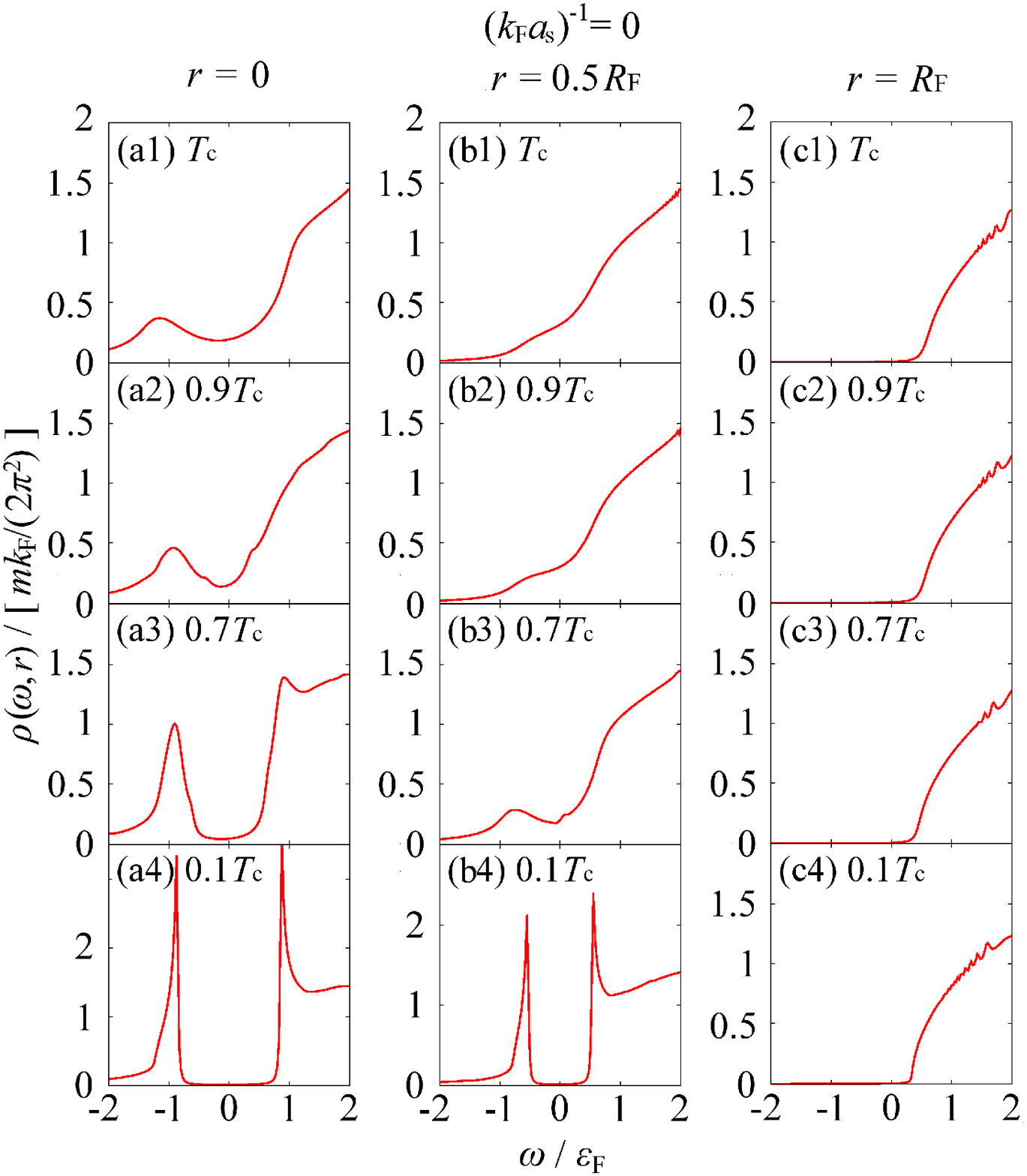}
\caption{(Color online) Calculated local density of states $\rho(\omega,r)$ in the unitarity limit ($(k_{\rm F}a_{\rm s})^{-1}=0$). (a1)-(a4) $r=0$. (b1)-(b4) $r=0.5R_{\rm F}$. (c1)-(c4) $r=R_{\rm F}$.}
\label{fig5}
\end{figure}
\par
\section{Pseudogap and superfluid gap in a trapped Fermi superfluid}
\par
Figure \ref{fig5} shows the local density of states (LDOS) $\rho(\omega,r)$ in the unitarity limit of a superfluid Fermi gas. In the trap center (panels (a1)-(a4)), a large dip structure associated with strong-pairing fluctuations is seen around $\omega=0$ at $T_{\rm c}$.
This pseudogap has already appeared above $T_{\rm c}$\cite{TSUCHIYA3} (although we do not show the result here).
At $0.9T_{\rm c}$ (panel (a2)), although the LDA superfluid order parameter $\Delta(r=0)$ is finite, LDOS still has a finite value at $\omega=0$, because of residual pairing fluctuations below $T_{\rm c}$.
The BCS-type full gap structure can be only seen below $T\simeq 0.8T_{\rm c}$.
Far below $T_{\rm c}$ (panel (a4)), the ordinary BCS-type density of states is realized, being accompanied by sharp coherence peaks at the excitation edges.
Panels (a1)-(a4) indicate that the pseudogapped LDOS at $r=0$ smoothly changes into the superfluid density of states with decreasing the temperature.
\par
At $r=0.5R_{\rm F}$, since pairing fluctuations are weaker than those in the trap center, the pseudogap (dip) structure does not appear at $T_{\rm c}$, as shown in Fig. \ref{fig5}(b1).
As mentioned previously, the LDA superfluid order parameter $\Delta(r)$ vanishes at $r\ge r_0(T)$, so that pairing fluctuations at $r>r_0$ continues to develop even below $T_{\rm c}$.
Because of this, the gradual development of the pseudogap around $\omega=0$ is seen in Fig. \ref{fig5}(b2) and (b3).
(We note that $\Delta(r=0.51R_{\rm F})=0$ at $T=0.7T_{\rm c}$.) Far below $T_{\rm c}$, since the LDA superfluid order parameter at $r=0.5R_{\rm F}$ becomes finite, the ordinary BCS superfluid density of states is obtained, as shown in Fig. \ref{fig5}(b4). 
\par
The particle density is very low around the edge of the gas cloud ($r\sim R_{\rm F}$), so that the superfluid order parameter is small and pairing fluctuations are weak there. As a result, LDOS at $r=R_{\rm F}$ shown in panels (c1)-(c4) is close to the density of states of a free Fermi gas,
\begin{equation}
\rho(\omega,r=R_{\rm F})\sim\sqrt{\omega+\mu(r)}\Theta(\omega+\mu(r)),
\end{equation}
where $\Theta(x)$ is the step function.
\par
\begin{figure}[t]
\includegraphics[width=.66\textwidth]{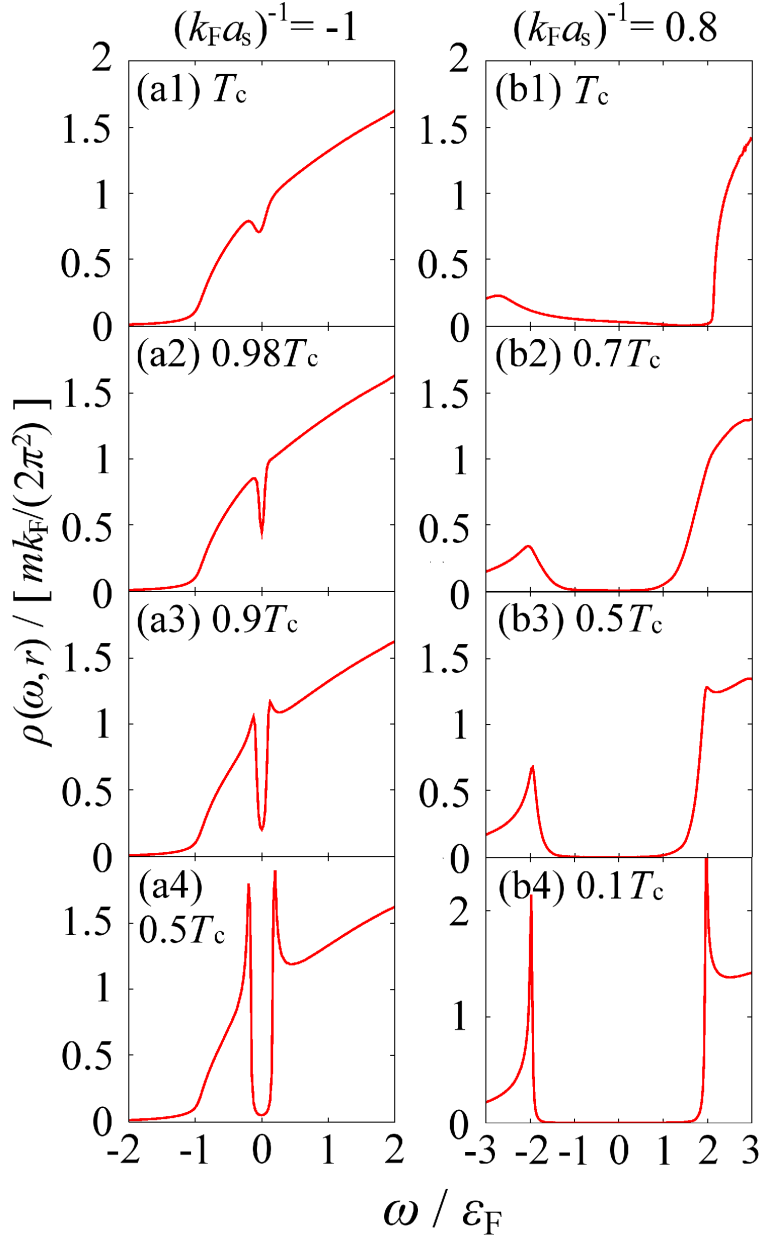}
\caption{(Color online) Calculated local density of state at $r=0$. (a1)-(a4) $(k_{\rm F}a_{\rm s})^{-1}=-1$ (BCS regime). (b1)-(b4) $(k_{\rm F}a_{\rm s})^{-1}=0.8$ (BEC regime).}
\label{fig6}
\end{figure}
\par
Since pairing fluctuations are weak in the BCS regime, the ordinary BCS-type superfluid density of states is soon realized below $T_{\rm c}$. (See Figs. \ref{fig6}(a1)-(a4).) On the other hand, LDOS in the BEC regime (Figs. \ref{fig6}(b1)-(b4)) is characterized by a fully gapped structure, reflecting the large binding energy of a tightly bound molecule, which has been already formed above $T_{\rm c}$. While the gap size is almost unchanged below $T_{\rm c}$, the growth of the coherence peaks can be seen with decreasing the temperature. 
\par
\begin{figure}[!t]
\includegraphics[width=12cm]{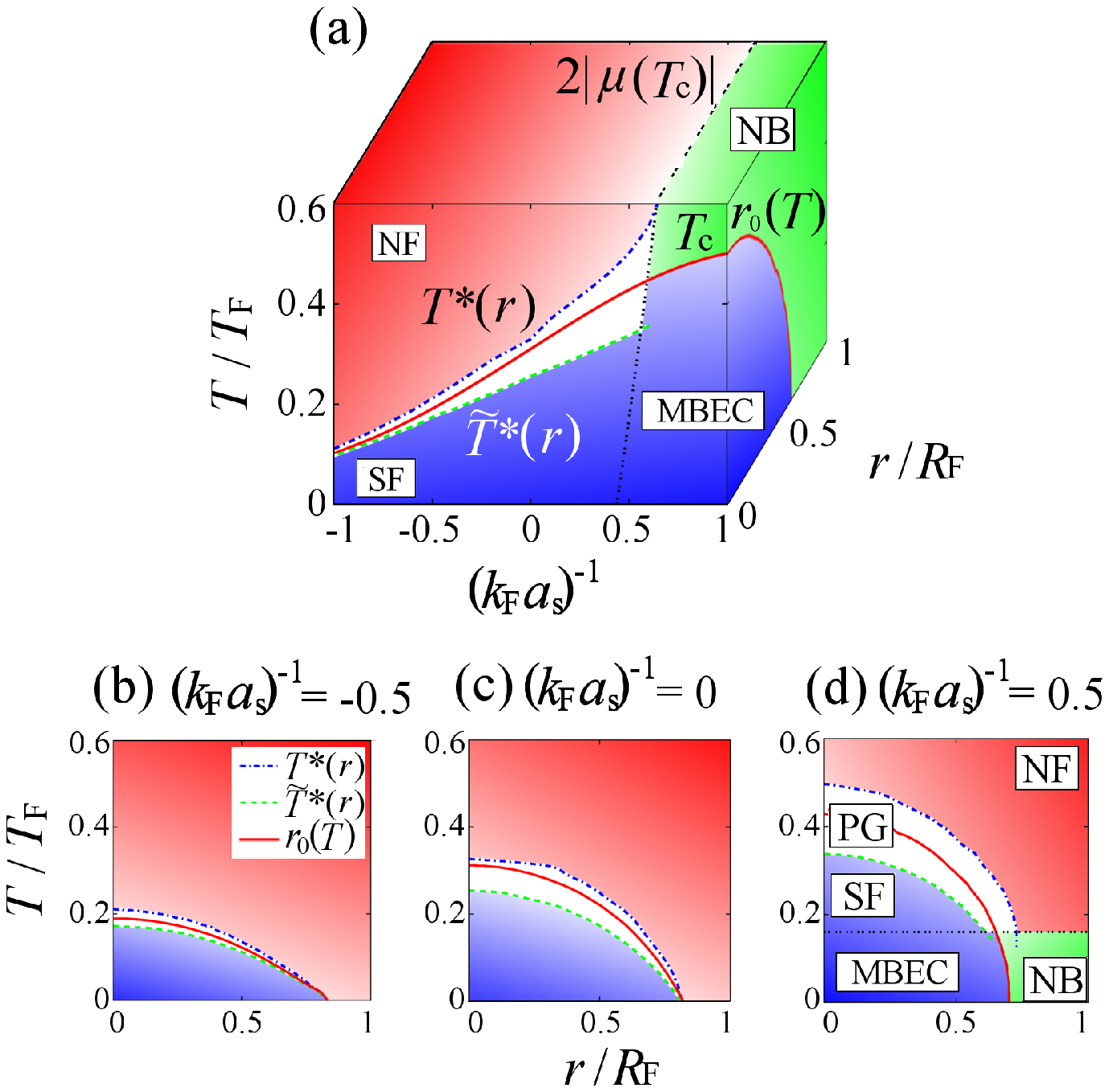}
\caption{(Color online) (a) Three-dimensional phase diagram of a trapped superfluid Fermi gas. The solid line at $r=0$ is $T_{\rm c}$. For the definitions of $T^*(r)$ and ${\tilde T}^*(r)$, see the text. $r_0(T)$ is the spatial position where the LDA superfluid order parameter vanishes at $T$. `SF' is the superfluid region where single-particle excitations are close to the ordinary BCS-type. `PG' is the pseudogap regime characterized by the pseudogapped local density of states. `NF' is the normal Fermi gas region where neither the superfluid gap nor the pseudogap appears in LDOS. In the BEC regime where $\mu<0$, we also plot $2|\mu(T_{\rm c})|$, which gives the characteristic temperature where two-body bound states appear. The right side of this line may be viewed as a molecular Bose gas, rather than a Fermi gas. In this regime, `MBEC' is the region which is well described by the BEC of tightly bound molecules. `NB' is the region of a non-condensed molecular Bose gas. We also show the tomographic views of this phase diagram in panels (b) $(k_{\rm F}a_{\rm s})^{-1}=-0.5$ (BCS side), (c) $(k_{\rm F}a_{\rm s})^{-1}=0$ (unitarity limit), and (d) $(k_{\rm F}a_{\rm s})^{-1}=0.5$ (BEC side).
}
\label{fig7}
\end{figure}

\begin{figure}[t]
\includegraphics[width=.66\textwidth]{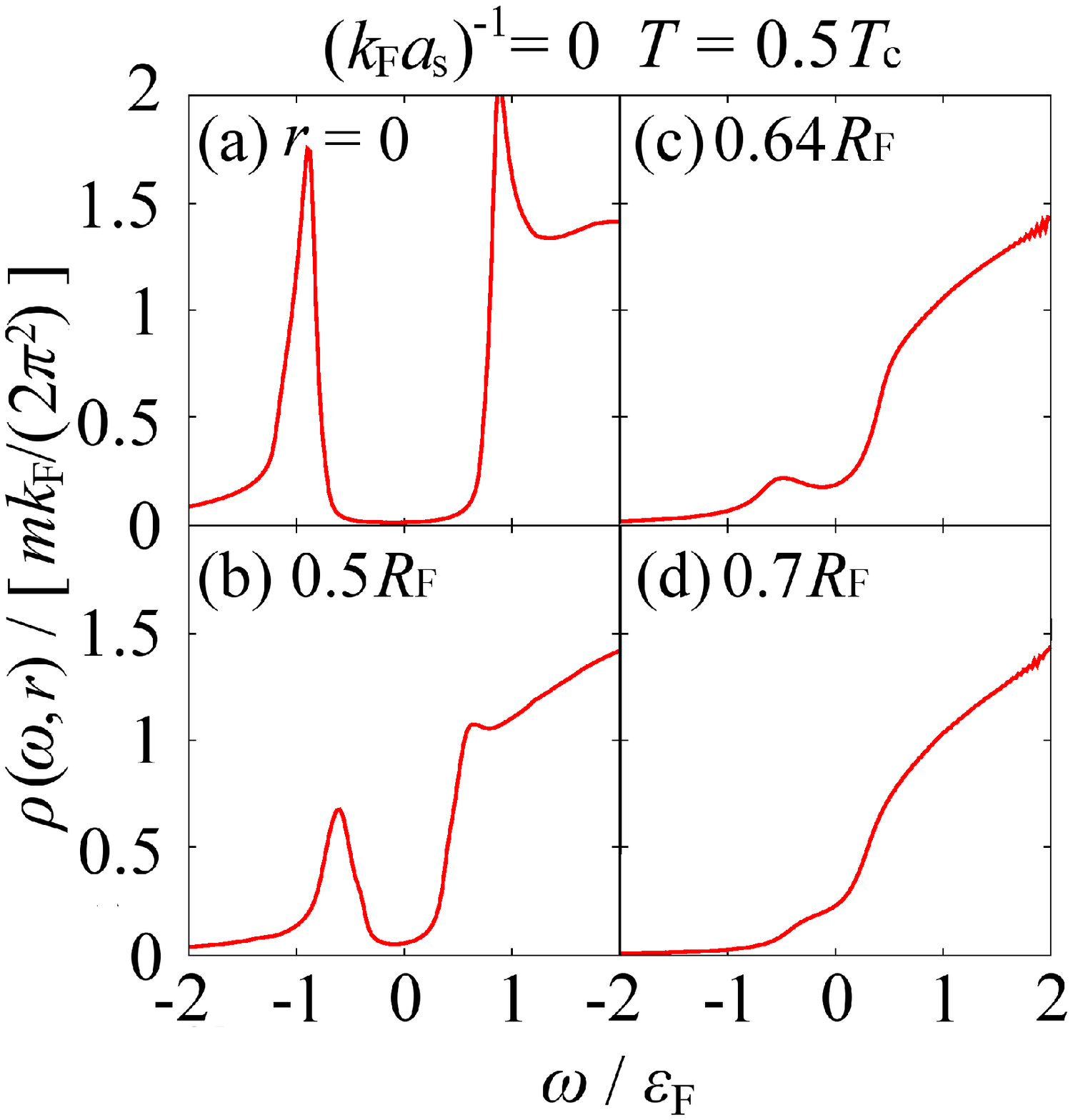}
\caption{(Color online) Spatial variation of LDOS at $T=0.5T_{\rm c}$ in the unitarity limit.}
\label{fig8}
\end{figure}
In our previous paper\cite{WATANABE1} for a uniform Fermi gas, we have introduced two characteristic temperatures $T^*$ and ${\tilde T}^*$ to identify the pseudogap regime. $T^*$ corresponds to the ordinary pseudogap temperature, namely, it is defined as the temperature at which a dip (pseudogap) structure appears in the density of states above $T_{\rm c}$. At ${\tilde T}^*$, the superfluid density of states $\rho(\omega=0)$ is suppressed by 50\% compared to the value of $\rho(\omega=0)$ at $T_{\rm c}$. Since the mean-field BCS state always has the vanishing density of states at $\omega=0$ below $T_{\rm c}$, the fact that $\rho(\omega=0)$ still has a large value at ${\tilde T}^*\le T\le T_{\rm c}$ means the importance of pairing fluctuations there. We have regarded the region ${\tilde T}^*\le T\le T^*$ as the pseudogap (PG) regime\cite{WATANABE1}, where pairing fluctuations dominate single-particle excitations. 
\par
Extending the above discussion to the present trapped case, we introduce two characteristic temperatures $T^*(r)$ and ${\tilde T}^*(r)$. $T^*(r)$ is defined as the temperature at which a dip (pseudogap) structure appears in $\rho(\omega,r)$ around $\omega=0$. ${\tilde T}^*(r)$ is determined from the condition that $\rho(\omega=0,r)$ is suppressed by 50\% compared with the value at the temperature where the LDA superfluid order parameter at $r$ becomes finite.
\par
Using $T^*(r)$ and ${\tilde T}^*(r)$, we obtain the phase diagram of a trapped Fermi gas in Fig. \ref{fig7}(a). At $r=0$, the overall structure is essentially the same as the phase diagram for a uniform Fermi gas\cite{WATANABE1}. As in the uniform case, we call the region between ${\tilde T}^*(r)$ and $T^*(r)$ the pseudogap (PG) region. The BCS-type superfluid density of states only appears below ${\tilde T}^*(r)$ (SF region). Above $T^*(r)$ (NF region), the pseudogap is absent in LDOS, where excitation properties are close to those of a normal Fermi gas.
\par
In Fig. \ref{fig7}(a), we also plot $2|\mu(T_{\rm c})|$ in the BEC regime where the Fermi chemical potential $\mu(T=T_{\rm c})$ is negative. Since $2|\mu|$ reduces to the binding energy $E_{\rm b}=1/(ma_s^2)$ of a two-body bound state in the BEC limit, this line physically describes the characteristic temperature where two-body bound molecules appear, overwhelming thermal dissociation. Thus, below $T\simeq 2|\mu(T_{\rm c})|$, the system may be viewed as a molecular Bose gas, rather than a Fermi atom gas. In this strong-coupling regime, $T_{\rm c}$ is well described by the BEC phase transition from a normal-state Bose gas (NB) to the molecular BEC (MBEC).
\par
We briefly note that, in Fig. \ref{fig7}, $T_{\rm c}$ is only the phase transition temperature. $T^*(r)$, ${\tilde T}^*(r)$, and $2|\mu(T_{\rm c})|$ are crossover temperatures without being accompanied by any phase transition.
\par
Figures \ref{fig7}(b)-(d) show that the PG region (${\tilde T}^*(r)\le T\le T^*(r)$) always exists along the $r_0(T)$-line determined by Eq. (\ref{eq.gap2}). Since the LDA order parameter $\Delta(r)$ becomes finite below this line, in a sense, the $r_0(T)$-line may be interpreted as the ``local superfluid phase transition temperature ($\equiv T_{\rm c}(r)$)" within the LDA picture\cite{NOTEZ}. Thus, these panels indicate that pairing fluctuations at $r$ become strong near $T_{\rm c}(r)$, leading to the pseudogap around $r_0(T)$.
\par
At a fixed temperature below $T_{\rm c}$, Figs. \ref{fig7}(b)-(d) indicate that a trapped Fermi gas exhibits a shell structure.
For example, in the unitarity limit at $T=0.5T_{\rm c}$, one finds from panel (c) that the SF, PG, and NF regions occupy the spatial regions, $0\le r<0.55R_{\rm F}$, $0.55R_{\rm F}\le r <0.67R_{\rm F}$, and $0.67R_{\rm F}<r$, respectively.
In this case, as shown in Fig. \ref{fig8}, while the BCS-type LDOS is obtained when $r\le 0.55R_{\rm F}$, the pseudogap is seen at $r=r_0=0.64R_{\rm F}$.
However, the pseudogap does not appear when $r\ge 0.67R_{\rm F}$. 
\par
As mentioned previously, the vanishing $\Delta(r)$ for $r\ge r_0$ is an artifact of LDA. In this regard, we note that the PG region also exists below the $r_0(T)$-lines in Figs. \ref{fig7}(b)-(d), where $\Delta(r)$ is small but finite. Thus, the PG region is expected to exist, even when one includes the finite value of $\Delta(r>r_0)$ by a more sophisticated treatment. In such an improved theory, since $\Delta(r>r_0)$ suppresses pairing fluctuations to some extent, the PG region would be narrower than the LDA result.  
\par
\begin{figure}[!t]
\includegraphics[width=5cm]{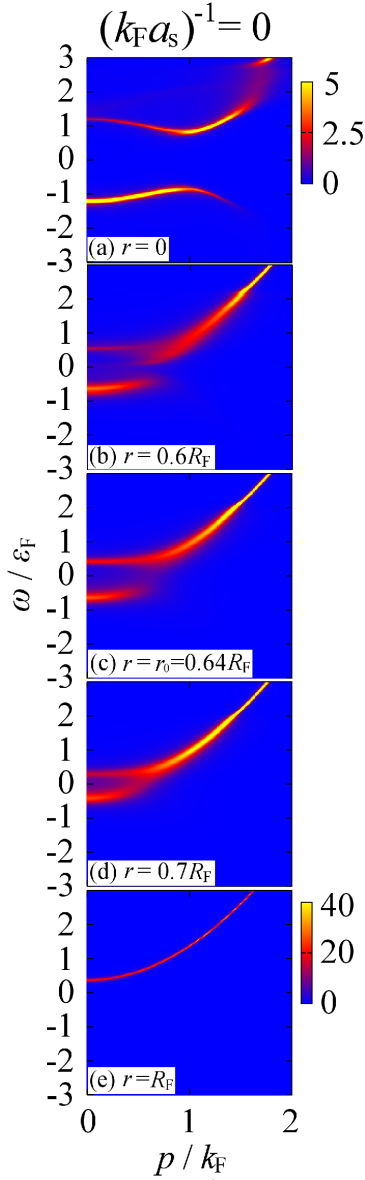}
\caption{(Color online) Intensity of local spectral weight (LSW) $A_{\bm p}(\omega,r)$ at $T=0.5T_{\rm c}$ in the unitarity limit. (a) $r=0$. (b) $r=0.6R_{\rm F}$. (c) $r=0.64R_{\rm F}~(=r_{\rm 0})$. (d) $r=0.7R_{\rm F}$. (e) $r=R_{\rm F}$. The intensity is normalized by $\varepsilon_{\rm F}^{-1}$. The same normalization is also used in Fig. \ref{fig10}.  
}
\label{fig9}
\end{figure}
\par
The ``SF-PG-NF shell structure" can be also seen in the spatial variation of local spectral weight (LSW) $A_{\bm p}(\omega,r)$ in the unitarity limit. In the trap center (Fig. \ref{fig9}(a)), the gapped spectral structure is close to the ordinary BCS-type spectral weight,
\begin{equation}
A_{\bm p}^{\rm BCS}(\omega,r=0)=
\sqrt{{1 \over 2}\left(1+{\xi_{\bm p} \over E_{\bm p}}\right)}
\delta(\omega-E_{\bm p})
+
\sqrt{{1 \over 2}\left(1-{\xi_{\bm p} \over E_{\bm p}}\right)}
\delta(\omega+E_{\bm p}),
\label{eq.bcssw}
\end{equation}
where $E_{\bm p}=\sqrt{\xi_{\bm p}+\Delta^2(0)}$ is the Bogoliubov single-particle excitation spectrum at $r=0$. Since $\Delta(r)$ is smaller in the outer region of the gas cloud, the gap size seen in LSW also becomes small, as shown in panel (b). However, although $\Delta(r)$ vanishes at $r=r_0=0.64R_{\rm F}$, $A_{\bm p}(\omega,r)$ in panel (c) still has a gap-like structure in the low momentum region, which is characteristic of the pseudogap phenomenon\cite{TSUCHIYA1,TSUCHIYA2}. This pseudogap becomes obscure in panel (d), to eventually disappear in panel (e). At $r=R_{\rm F}$, the spectral peak line is close to the free-particle dispersion $\omega=p^2/(2m)-\mu(R_{\rm F})$.
\par
\begin{figure}[!t]
\includegraphics[width=10cm]{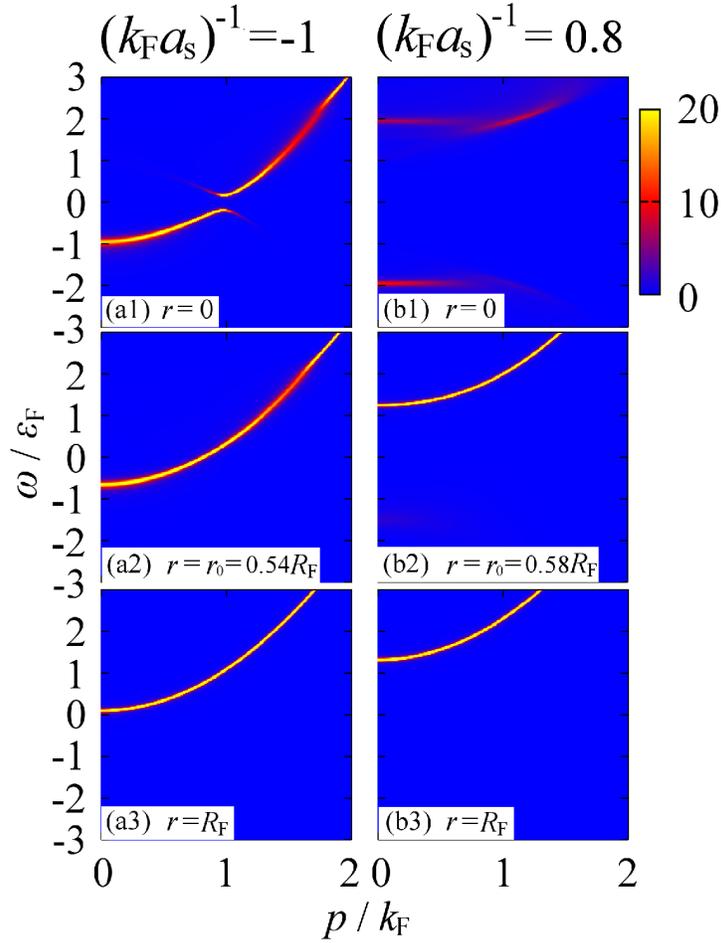}
\caption{(Color online) Intensity of LSW at $T=0.5T_{\rm c}$. (a1)-(a3)  $(k_{\rm F}a_{\rm s})^{-1}=-1$ (BCS regime). (b1)-(b3) $(k_{\rm F}a_{\rm s})^{-1}=0.8$ (BEC regime).
}
\label{fig10}
\end{figure}
\par
In the weak-coupling BCS regime, while the BCS-type gapped spectrum, as well as the free-particle-like peak line, can be seen at $r=0$ and $r=R_{\rm F}$, respectively, a pseudogap structure does not appear at $r=r_0$, as shown in Fig. \ref{fig10}(a1)-(a3). In the uniform case, it has been shown\cite{TSUCHIYA1} that the pseudogap in the BCS regime can be more clearly seen in the density of states than the spectral weight, leading to different pseudogap temperatures that are determined from these quantities. In particular, when $(k_{\rm F}a_s)^{-1}=-1$, the pseudogap is almost invisible in the spectral weight even at $T_{\rm c}$\cite{TSUCHIYA1}, although a dip structure appears in the density of states. In the present trapped case, a similar situation is considered to occur in panel (a2), although the pseudogap region exists along the $r_0$-line in the phase diagram in Fig. \ref{fig7} (which is obtained from the local density of states).
\par
In the BEC regime with $\mu<0$, Figs. \ref{fig10}(b1)-(b3) show that, while the lower peak line gradually disappears with increasing $r$, the upper branch reduces to $\omega=p^2/(2m)-\mu(r)$ around the edge of the gap cloud. The upper branch is related to the dissociation of tightly bound molecules that have been already formed above $T_{\rm c}$, so that it exists even for $\Delta(r)=0$. On other hand, since the lower branch is associated with a particle-hole coupling induced by the superfluid order parameter\cite{TSUCHIYA1,TSUCHIYA2}, it is absent when $r\ge r_0(T)$ (where $\Delta(r)=0$).
\par
\begin{figure}[!t]
\includegraphics[width=0.66\textwidth]{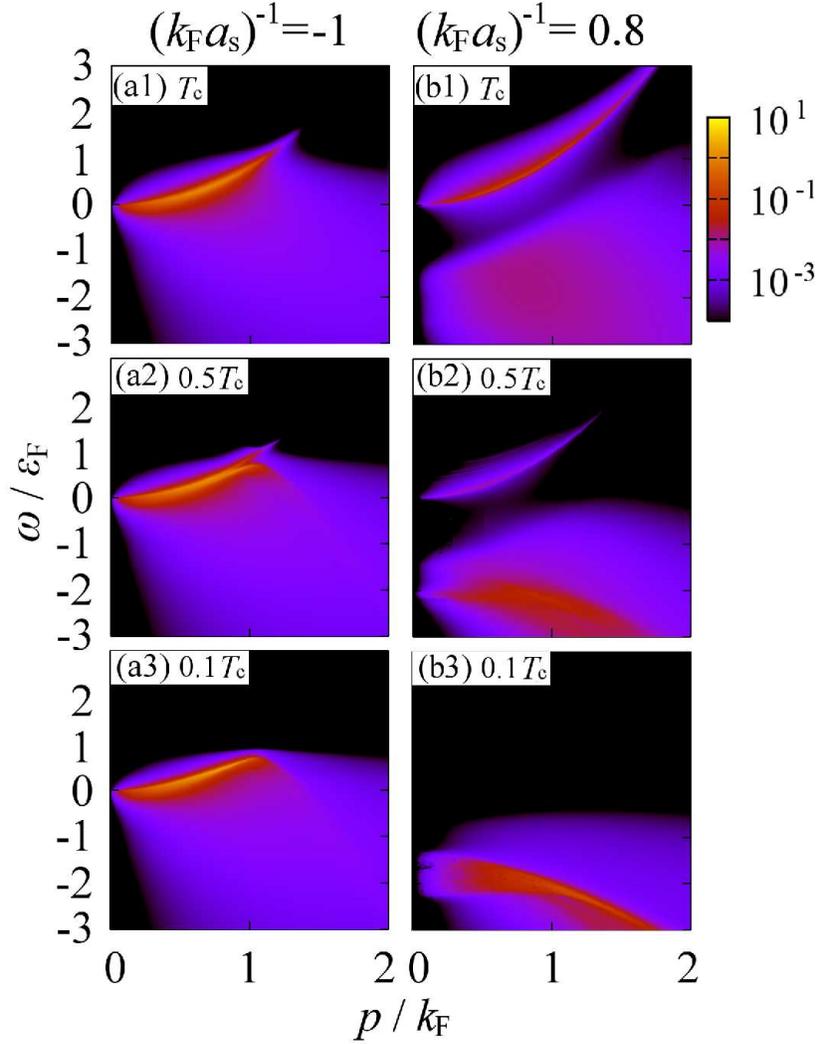}
\caption{(Color online) Calculated intensity of photoemission spectrum $p^2\overline{A_{\bm p}(\omega)f(\omega)}$. (a1)-(a3) $(k_Fa_s)^{-1}=-1$ (BCS regime). (b1)-(b3) $(k_Fa_s)^{-1}=0.8$ (BEC regime). The intensity is normalized by $2\pi t_{\rm F}^2/(2m)$. The same normalization is used in Fig. \ref{fig12}
}
\label{fig11}
\end{figure}
\par
\begin{figure}[!t]
\includegraphics[width=0.33\textwidth]{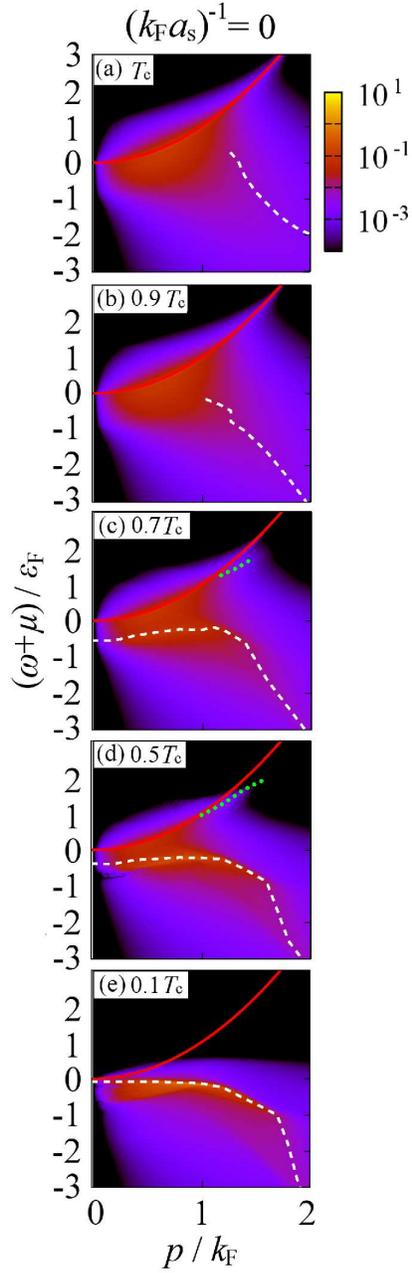}
\caption{(Color online) Calculated intensity of photoemission spectrum $p^2\overline{A_{\bm p}(\omega)f(\omega)}$ in the unitarity limit $(k_{\rm F}a_s)^{-1}=0$. The solid line is the free-particle dispersion $\omega+\mu=p^2/(2m)$. The dashed line and dotted line show peak positions of the spectrum.
}
\label{fig12}
\end{figure}
\par
\section{Photoemission spectrum in a trapped superfluid Fermi gas}
\par
Figures \ref{fig11} and \ref{fig12} show photoemission spectra $p^2\overline{A_{\bm p}(\omega)f(\omega)}$. As discussed in Ref.\cite{TSUCHIYA2}, the calculated spectra at $T_{\rm c}$ (Figs. \ref{fig11}(a1), 11(b1) and \ref{fig12}(a)) agree well with the recent experiment on a $^{40}$K Fermi gas\cite{STEWART,GAEBLER}. That is, starting from the weak-coupling BCS regime, a sharp peak line along the free particle dispersion $\omega+\mu=p^2/(2m)$ in Fig. \ref{fig11}(a1) becomes broad in the unitarity limit (Fig. \ref{fig12}(a1)), which eventually splits into an upper sharp branch and a lower broad branch in the BEC regime (Fig. \ref{fig11}(b1)).
\par
In the weak-coupling BCS regime, the overall spectral structure almost remains unchanged below $T_{\rm c}$, as shown in Fig. \ref{fig11}(a1)-(a3). However, when we carefully look at panel (a2), we find the splitting of the spectral peak around $p/k_{\rm F}=1$. The upper peak is along the free particle dispersion $\omega+\mu=p^2/(2m)$, so that it comes from the NF region around the edge of the gas cloud. On the other hand, the lower peak line does not exist in panel (a1), and the momentum dependence is similar to the hole branch of the BCS Bogoliubov excitation spectrum (which is given by $E_{\bm p}=-\sqrt{(p^2/(2m)-\varepsilon_{\rm F})^2+\Delta^2}$ in the BCS theory). Thus, the lower peak is considered to come from the SF region around the trap center. In this sense, the spectral structure seen in Fig. \ref{fig11}(a2) is consistent with the shell structure discussed in Fig. \ref{fig7}. Since the SF region spreads out to the whole the gas cloud far below $T_{\rm c}$, the upper peak line disappears in Fig. \ref{fig11}(a3).
\par
In the unitarity limit at $T_{\rm c}$, while the peak line along the free particle dispersion $\omega+\mu=p^2/(2m)$ in Fig. \ref{fig12}(a) comes from the NF region, the broad spectral structure below this reflects the PG region\cite{TSUCHIYA2,TSUCHIYA3}. Since the SF region only appears below ${\tilde T}^*(r=0)=0.81T_{\rm c}$ (See the phase diagram in Fig. \ref{fig7}.), the photoemission spectrum almost remains unchanged at $T=0.9T_{\rm c}$, as shown in Fig. \ref{fig12}(b). The SF region starts to develop from the trap center below ${\tilde T}^*(r=0)$, so that the spectral weight gradually move to the lower peak line, as shown in panels (c)-(e).
In this low temperature region, in addition to the hole branch of Bogoliubov excitations, one also slightly sees the peak line corresponding to the particle branch of Bogoliubov excitations below the free particle dispersion (dotted line in panels (c) and (d).)
Far below $T_{\rm c}$, because the SF region covers the whole gas cloud, and because the Fermi distribution function in Eq. (\ref{eq.101}) suppresses the spectral intensity in the high energy region, the photoemission spectrum is dominated by the lower Bogoliubov branch, as shown in Fig. \ref{fig12}(e). 
\par
In the strong-coupling BEC regime, Figs. \ref{fig11}(b1)-(b3) show that the photoemission spectrum already splits into an upper and lower branches at $T_{\rm c}$, reflecting a large molecular binding energy. Since the upper branch is suppressed by the Fermi distribution function in Eq. (\ref{eq.101}) far below $T_{\rm c}$, the lower branch is only seen in Fig. \ref{fig11}(b3).
\par
\begin{figure}[!t]
\includegraphics[width=6cm]{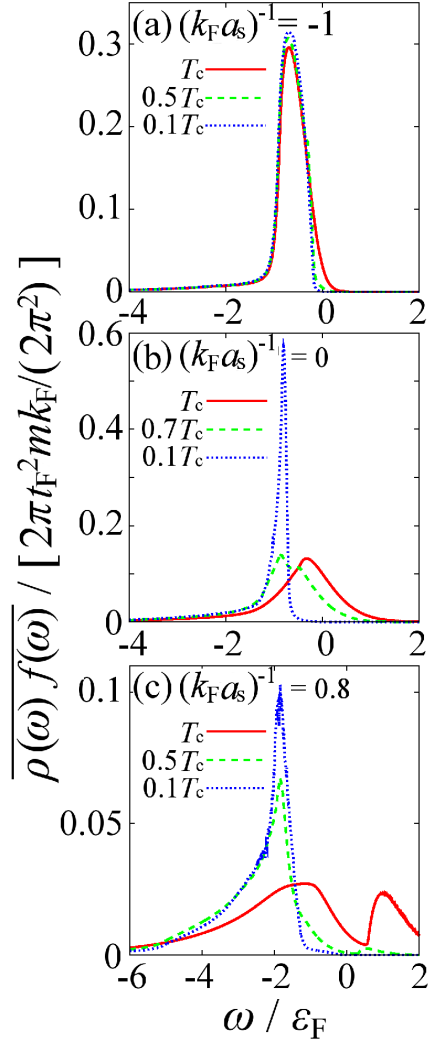}
\caption{(Color online) Calculated occupied density of states $\overline{\rho(\omega)f(\omega)}$ in the superfluid state below $T_{\rm c}$. (a) $(k_{\rm F}a_{\rm s})^{-1}=-1$ (BCS regime). (b) $(k_{\rm F}a_{\rm s})^{-1}=0$ (unitarity limit). (c) $(k_{\rm F}a_{\rm s})^{-1}=-0.8$ (BEC regime).
}
\label{fig13}
\end{figure}
\par
\begin{figure}[!t]
\includegraphics[width=12cm]{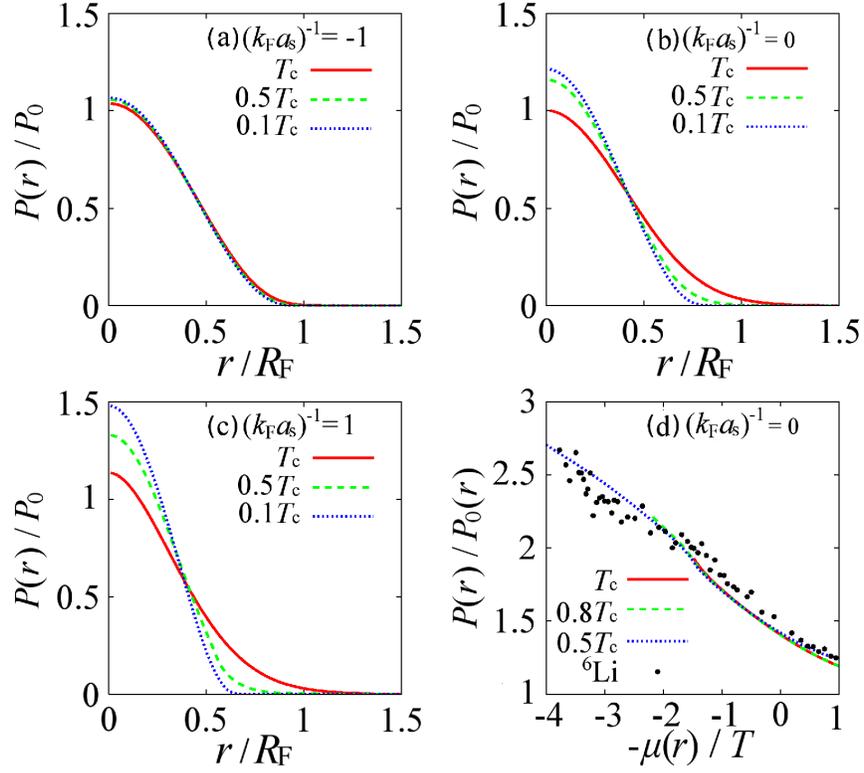}
\caption{(Color online) Calculated local pressure $P(r)$ in the BCS-BEC crossover regime of a trapped Fermi gas. (a) $(k_{\rm F}a_{\rm s})^{-1}=-1$ (BCS regime). (b) $(k_{\rm F}a_{\rm s})^{-1}=0$ (unitarity limit). (c) $(k_{\rm F}a_{\rm s})^{-1}=1$ (BEC regime). In panel (d), we plot $P(r)$ as a function of $\mu(r)/T$. Experimental results on a $^6$Li Fermi gas\cite{Nascimbene} are also shown in panel (d) (solid circles). $P_0=\frac{1}{15\pi^2}(2m)^{3/2}\epsilon_{\rm F}^{5/2}$ is local pressure of a free Fermi gas at $r=0$ and $T=0$. $P_0(r)=m\omega^2\int_{\infty}^r r^\prime dr^\prime\sum_{\bm p}f(\xi_p(r^\prime))$ is the local pressure of a free Fermi gas where the same values of $T$ and $\mu$ as those used in calculating $P(r)$ are taken.
}
\label{fig14}
\end{figure}

Figure \ref{fig13} shows the occupied density of states $\overline{\rho(\omega)f(\omega)}$ in a trapped superfluid Fermi gas. In the BCS regime, as expected from the weak temperature dependence of the photoemission spectrum in Figs. \ref{fig11}(a1)-(a3), the occupied density of states $\overline{\rho(\omega)f(\omega)}$ in Fig. \ref{fig14}(a) almost remains unchanged below $T_{\rm c}$. In contrast, in the unitarity limit (panel (b)), a peak starts to grow at $\omega/\varepsilon_{\rm F}\simeq -1$ below $T\lesssim 0.7T_{\rm c}$ to become a sharp peak far below $T_{\rm c}$. From the comparison with Fig. \ref{fig12}(c), we find that this peak corresponds to the hole branch of Bogoliubov single-particle excitations. The growth of this sharp peak can be also seen in the BEC regime, as shown in Fig. \ref{fig13}(c). 
\par
\par
\section{Local pressure in a trapped Fermi gas}
\par
Figures \ref{fig14}(a)-(c) show the local pressure $P(r)$ in the BCS-BEC crossover regime of a trapped superfluid Fermi gas. In the BCS regime, panel (a) shows that $P(r)$ is almost $T$-independent. Since $P(r)$ is related to the particle density $n(r)$ as Eq. (\ref{eq.P})\cite{NOTEP}, this result reflects the weak temperature dependence of $n(r)$ in this region. (See Fig. \ref{fig3}(a)). When the interaction strength becomes strong, Figs. \ref{fig3}(b) and (c) indicate that particles tend to cluster around the trap center below $T_{\rm c}$. This enhances the local pressure $P(r)$ around $r=0$, as well as the decrease of $P(r)$ around $r=R_{\rm F}$, below $T_{\rm c}$, as shown in Fig. \ref{fig14}(b) and (c). 
\par
In the unitarity limit, since $a_{\rm s}$ diverges, the system can be described by the single dimensionless parameter $\mu(r)/T$\cite{Ho}. This university also holds in the present combined $T$-matrix theory with LDA, as discussed in the Appendix. Indeed, when we plot $P(r)$ as a function of $\mu(r)/T$, all the results at different temperatures are well fitted by a universal function, as shown in Fig. \ref{fig14}(d).
\par
In panel (d), we also compare our result with the recent experiment on a $^6$Li Fermi gas done by ENS group\cite{Nascimbene}. Without introducing any fitting parameter, our result is in good agreement with the experiment.
\par
We note that, although our theory correctly include the pseudogap effect associated with strong pairing fluctuations, one cannot see a clear signature of this strong-coupling effect in Fig. \ref{fig14}. Indeed, Ref.\cite{Nascimbene} reports that their experimental data can be well described by the Fermi liquid theory. From these, one finds that the local pressure $P(r)$ is not sensitive to  the pseudogap phenomenon. The LDA particle density $n(r)$ in Eq. (\ref{eq.P}) is given by
\begin{equation}
n(r)=\int_{-\infty}^{\infty}d\omega f(\omega)\rho(\omega,r).
\label{AAA}
\end{equation}
Thus, even when the LDOS $\rho(\omega,r)$ has a pseudogap structure around $\omega=0$, it would be smeared out to some extent by the $\omega$-integration in Eq. (\ref{AAA}). The pseudogap effect would be further smeared out by the spatial integration in Eq. (\ref{eq.P}). Because of these two integrations, the detailed pseudogap structure in the low energy density of states is considered to be not crucial for $P(r)$.
\par
\section{Summary}
\par
To summarize, we have discussed pseudogap phenomena and effects of a harmonic trap in the BCS-BEC crossover regime of a superfluid Fermi gas. Extending our previous work for a uniform system to include effects of a harmonic trap within the local density approximation (LDA), we clarified the coexistence of the superfluid (SF) region where the BCS-type superfluid density of states appears and the pseudogap (PG) region which is dominated by pseudogap phenomenon even below $T_{\rm c}$. From the spatial and temperature dependence of the local density of states, we identified the pseudogap region in the phase diagram with respect to the temperature, interaction strength, and spatial position.
\par
We have discussed the photoemission spectrum in a trapped superfluid Fermi gas. In the BCS-BEC crossover region, the photoemission spectrum was shown to be strongly affected by the shell structure of a trapped superfluid Fermi gas which consists of the SF region, PG region, and the normal Fermi gas (NF) region. Since the inhomogeneity and strong-pairing fluctuations are important key issues in considering a real trapped Fermi gas, our results would be useful for the study of strong-coupling effects of this system, including the realistic situation. 
\par
We have also examined pseudogap effects on the local pressure $P(r)$. While our strong-coupling result agrees well the recent experiment on a $^6$Li Fermi gas done by ENS group\cite{Nascimbene}, we showed that this thermodynamic quantity is not sensitive to the pseudogap appearing in the single-particle density of states. This is consistent with the statement\cite{Nascimbene} that the observed pressure in the unitarity limit can be well described by the Fermi liquid theory. Since the pressure is not directly related to the detailed single-particle excitations compared with the photoemission spectrum, it is possible to occur that, while the pseudogap can be observed in the latter, such a strong-coupling phenomenon does not clearly appear in the former thermodynamic quantity. 
\par
In this paper, we have treated the inhomogeneity of the system within LDA. While LDA has succeeded in explaining various properties of trapped Fermi gases, it cannot correctly describe the feature that the superfluid order parameter is finite everywhere below $T_{\rm c}$. To overcome this, one needs a more sophisticated inhomogeneous strong-coupling theory than LDA. Since the presence of a trap potential is unique to the cold Fermi gas system, this problem would be an important challenge for the further development of the BCS-BEC crossover theory in cold atom physics.  
\par
\acknowledgments
\par
We would like to thank T. Kashimura, S. Watabe Y. Endo, D. Inotani and R. Hanai for fruitful discussions.
R.W. was supported by the Japan Society for the Promotion of Science. Y. O. was supported by Grant-in-Aid for Scientific research from MEXT in Japan (No.22540412, No.23104723, No.23500056).
\par
\par
\appendix
\section{Universality within the $T$-matrix approximation}
\par
To see the universality of the present combined $T$-matrix theory with LDA, it is convenient to write the dimensionless self-energy $\tilde{\Sigma}_{\bm \tilde{p}}(i{\tilde \omega}_n,\tilde{a}_s,\tilde{\mu}(r),\tilde{\Delta}(r))\equiv T^{-1}\Sigma_{\bm p}(i\omega_n,r)$ in the present approximation in the form
\begin{eqnarray}
\tilde{\Sigma}_{\bm \tilde{p}}(i{\tilde \omega}_n,\tilde{a}_s,\tilde{\mu}(r),\tilde{\Delta}(r))
&=&-T\sum_{{\tilde \nu}_{n'}}\sum_{ss^\prime=\pm}\int_0^\infty \tilde{q}^2d\tilde{q}\sin\theta d\theta\tilde{\Gamma}^{ss^\prime}_{\bm \tilde{q}}(i{\tilde \nu}_{n'},r)\nonumber\\
&&\qquad\times\tau_{-s}\frac{1}{i{\tilde \omega}_{n+n'}-\tilde{\xi}_{\bm {q+p}}(r)\tau_3+\tilde{\Delta}(r)\tau_1} \tau_{-s^\prime},
\label{eq.A1}
\end{eqnarray}
where $\tilde{p}=\sqrt{\varepsilon_p/T}$, $\tilde{q}=\sqrt{\varepsilon_{\bm q}/T}$, $\cos\theta=\tilde{{\bm p}}\cdot\tilde{{\bm q}}/(\tilde{p}\tilde{q})$, $\tilde{\mu}(r)=\mu(r)/T$, $\tilde{\Delta}(r)=\Delta(r)/T$, $\tilde{\xi}_{\bm p}=\xi_{\bm p}/T$, and ${\tilde a}_s=\sqrt{2mT}a_s$. In Eq. (\ref{eq.A1}), the dimensionless fermion Matsubara frequency ${\tilde \omega}_n=\omega_n/T=(2n+1)\pi$ no longer has any physical quantity.
In the same manner, the dimensionless boson Matsubara frequency is simply given by ${\tilde \nu}_{n}=\nu_{n}/T=2n\pi$.
The particle-particle scattering matrix ${\tilde \Gamma}^{ss^\prime}_{{\tilde {\bm q}}}(i{\tilde \nu}_{n},r)\equiv(2\pi)^{-2}\sqrt{(2m)^3T}\Gamma^{ss^\prime}_{\bm q}(i\nu_n,r)$ in Eq. (\ref{eq.A1}) has the form,
\begin{equation}
\left(
\begin{array}{cc}
{\tilde \Gamma}^{+-}_{\bm q}(i{\tilde \nu}_{n},r)&{\tilde \Gamma}^{++}_{\bm q}(i{\tilde \nu}_{n},r)\\
{\tilde \Gamma}^{--}_{\bm q}(i{\tilde \nu}_{n},r)&{\tilde \Gamma}^{-+}_{\bm q}(i{\tilde \nu}_{n},r)
\end{array}
\right)=\left[\frac{\pi}{2}\frac{1}{\tilde{a}_{\rm s}}-\int_0^\infty d\tilde{p}-\left(
\begin{array}{cc}
\tilde{\Pi}_{\bm \tilde{q}}^{-+}(i{\tilde \nu}_{n},\tilde{\mu}(r),\tilde{\Delta}(r))&\tilde{\Pi}_{\bm \tilde{q}}^{++}(i{\tilde \nu}_{n},\tilde{\mu}(r),\tilde{\Delta}(r))\\
\tilde{\Pi}_{\bm \tilde{q}}^{--}(i{\tilde \nu}_{n},\tilde{\mu}(r),\tilde{\Delta}(r))&\tilde{\Pi}_{\bm \tilde{q}}^{+-}(i{\tilde \nu}_{n},\tilde{\mu}(r),\tilde{\Delta}(r))
\end{array}
\right)
\right]^{-1}.
\label{BBB}
\end{equation}
Here, the second term on the right hand side of Eq. (\ref{BBB}) has been introduced to regularize the theory.
The correlation function $\tilde{\Pi}_{\tilde{q}}^{ss^\prime}(i{\tilde \nu}_{n},\tilde{\mu}(r),\tilde{\Delta}(r))\equiv ((2\pi)^2/\sqrt{(2m)^3T})\times \Pi_{\bm q}^{ss^\prime}(i\nu_{n},r)$ is given by, for example, 
\begin{eqnarray}
\tilde{\Pi}_{\bm \tilde{q}}^{++}
(i{\tilde \nu}_{n},\tilde{\mu}(r),\tilde{\Delta}(r))
&=&
\frac{1}{4}\sum_{s=\pm1}\int_0^\infty \tilde{p}^2d\tilde{p}\sin\theta d\theta
 \frac{s\tilde{\Delta}(r)^2}
 {\tilde{E}_{{\bm p}+{\bm q}/2}(r)\tilde{E}_{{\bm p}-{\bm q}/2}(r)}\nonumber\\
&&\qquad\times\frac{\tilde{E}_{{\bm p}+{\bm q}/2}(r)+s\tilde{E}_{{\bm p}-{\bm q}/2}(r)}
 {(2n\pi)^2+(\tilde{E}_{{\bm p}+{\bm q}/2}(r)+s\tilde{E}_{{\bm p}-{\bm q}/2}(r))^2}
\nonumber\\ 
&&\qquad\times\left[\tanh\left({\tilde{E}_{{\bm p}+{\bm q}/2}(r)\over 2}\right)+s\tanh\left({\tilde{E}_{{\bm p}-{\bm q}/2}(r) \over 2}\right)\right],
\end{eqnarray}
where $\tilde{E}_{\bm p}(r)=E_{\bm p}(r)/T=\sqrt{(\tilde{p}^2-\tilde{\mu}(r))^2+\tilde{\Delta}(r)^2}$.
\par
The dimensionless LDA superfluid order parameter $\tilde{\Delta}(r)$ obeys the gap equation, 
\begin{equation}
1=-\frac{2}{\pi}\tilde{a}_{\rm s}\int_0^{\infty}\tilde{p}^2d\tilde{p}\left(\frac{1}{\tilde{E}_{\bm p}(r)}\tanh\frac{\tilde{E}_{\bm p}(r)}{2}-1\right).
\end{equation}
Thus, $\tilde{\Delta}(r)$ is found to be a function of $({\tilde a}_s,{\tilde \mu}(r))$.
\par
Using Eq. (\ref{eq.A1}), we find that the dimensionless Green's function $\tilde{G}_{\bm \tilde{p}}(i{\tilde \omega}_n,\tilde{a}_{\rm s},\tilde{\mu}(r))\equiv TG_{\bm p}(i\omega_n,r)$ only depends on ${\tilde a}_{\rm s}$ and $\tilde{\mu}(r)$, as
\begin{equation}
\tilde{G}_{\bm \tilde{p}}(i{\tilde \omega}_n,\tilde{a}_s,\tilde{\mu}(r))=\frac{1}{i{\tilde \omega}_n-\tilde{\xi}_{\bm {\tilde{p}}}(r)\tau_3+\tilde{\Delta}(\tilde{a}_{\rm s},\tilde{\mu}(r))\tau_1-\tilde{\Sigma}_{\bm \tilde{p}}(n,\tilde{a}_{\rm s},\tilde{\mu}(r))}.
\label{eq.CCC}
\end{equation}
We note that ${\tilde \omega}_n=(2n+1)\pi$ does not involve a physical quantity. In the analytic-continued form of Eq. (\ref{eq.CCC}) is given by
\begin{equation}
\tilde{G}_{\bm \tilde{p}}(\tilde{\omega}+i\delta,\tilde{a}_s,\tilde{\mu}(r))=\frac{1}{\tilde{\omega}+i\delta-\tilde{\xi}_{\bm {\tilde{p}}}(r)\tau_3+\tilde{\Delta}(\tilde{a}_{\rm s},\tilde{\mu}(r))\tau_1-\tilde{\Sigma}_{\bm \tilde{p}}(\tilde{\omega}+i\delta,\tilde{a}_{\rm s},\tilde{\mu}(r))},
\end{equation}
where $\tilde{\omega}=\omega/T$. As a result, any physical quantity calculated from $G_{\bm p}(i\omega_n,r)$ can be written in the form $AT^\alpha F(\tilde{a}_{\rm s},\tilde{\mu}(r))$, where the coefficient $A$ and the exponent $\alpha$ depend on the detailed physical quantity we are considering. $F(\tilde{a}_{\rm s},\tilde{\mu}(r))$ is a dimensionless function calculated from ${\tilde G}$, which only depends on $(\tilde{a}_{\rm s},\tilde{\mu}(r))$. Thus, when the physical quantity is normalized by $AT^\alpha$, it exhibits a universal behavior with respect to $(\tilde{a}_{\rm s},\tilde{\mu}(r))$. In particular, in the unitarity limit (where the scattering length $a_s$ diverges), the universal behavior is dominated by the single parameter $\mu(r)/T$.
\par
\par


\begin{thebibliography}{99}
\bibitem{noteZ} We actually subtract $\langle\rho_1({\bm q}=0)\rangle$ from the $\tau_1$-component $\rho_1({\bm q})$, although we do not explicitly write this prescription in the definition of $\rho_1({\bm q})$.
\bibitem{NOTEZ} Since the superfluid phase transition is a bulk phenomenon, $T_{\rm c}$ must be unique. We only use the ``local superfluid phase transition temperature $T_{\rm c}(r)$" to simply explain the reason why the pseudogap appears around the $r_0(T)$-line.
\bibitem{NOTEP} Although $P(r)$ is deeply related to the LDA density profile $n(r)$, the former is obtained as a smooth function of $r$. This is in contrast to the latter where a cusp structure appears at $r_0(T)$, as shown in Fig. \ref{fig3}.
\end{thebibliography}
\end{document}